\documentclass[runningheads,a4paper]{llncs}
\usepackage[english]{babel}
\usepackage[utf8]{inputenc}
\usepackage{csquotes}
\usepackage{cite}
\usepackage{hyperref}
\usepackage{xspace}
\usepackage{subcaption}
\usepackage{centernot}
\usepackage{color}
\usepackage{graphicx}
\usepackage[binary-units]{siunitx}
\usepackage{xcolor}
\usepackage{xparse}
\usepackage{amsmath}
\usepackage{amssymb}
\usepackage{listings-rust}
\usepackage{listings}
\usepackage{dsfont}
\usepackage[inline]{enumitem}
\usepackage{subcaption}
\usepackage{tikz}
\usepackage[color=blue!20, linecolor=blue!80!black]{todonotes}

\usetikzlibrary{decorations.pathreplacing}
\usetikzlibrary{arrows,automata,positioning,shapes,patterns,arrows.meta}

\usepackage[capitalise,nameinlink]{cleveref}
\crefname{section}{Sect.}{Sect.}
\Crefname{section}{Section}{Sections}

\newcommand{\donotshow}[1]{}

\definecolor{CommentColor}{rgb}{0.13, 0.13, 1}

\definecolor{first}{HTML}{70ACEC}
\definecolor{second}{HTML}{5682B1}
\definecolor{third}{HTML}{3B5979}

\definecolor{time}{HTML}{00A4CC}
\definecolor{mem}{HTML}{F95700}

\definecolor{prelude}{HTML}{F79A3D}
\definecolor{loop}{HTML}{2874ae}
\definecolor{prefix}{HTML}{61B940}
\definecolor{postfix}{HTML}{A370AB}

\newcommand*\hlc{\textsc{hlc}\xspace}
\newcommand*\llc{\textsc{llc}\xspace}

\DeclareFontFamily{U}{MnSymbolC}{}
\DeclareSymbolFont{MnSyC}{U}{MnSymbolC}{m}{n}
\DeclareFontShape{U}{MnSymbolC}{m}{n}{
    <-6>  MnSymbolC5
   <6-7>  MnSymbolC6
   <7-8>  MnSymbolC7
   <8-9>  MnSymbolC8
   <9-10> MnSymbolC9
  <10-12> MnSymbolC10
  <12->   MnSymbolC12%
}{}
\DeclareMathSymbol{\powerset}{\mathord}{MnSyC}{180}

\newcommand*{\eg}{e.g.\@\xspace}
\newcommand*{\ie}{i.e.\@\xspace}

\newcommand*{\wrt}{w.r.t.\@\xspace}
\makeatletter
\newcommand*{\etc}{%
  \@ifnextchar{.}%
    {etc}%
    {etc.\@\xspace}%
}
\newcommand*{\cf}{%
  \@ifnextchar{.}%
    {cf}%
    {cf.\@\xspace}%
}
\newcommand*{\etal}{%
  \@ifnextchar{.}%
    {et~al}%
    {et~al.\@\xspace}%
}
\makeatother

\NewDocumentCommand{\twopartdef}{ m m m o }%
{
  \left\{
    \begin{array}{ll}
      #1\quad & \mbox{if } #2 \\
      #3\quad & \IfValueTF{#4}{\mbox{if } #4}{\mbox{otherwise}}
    \end{array}
  \right.
}
\NewDocumentCommand{\threepartdef}{ m m m m m o }%
{
  \left\{
    \begin{array}{ll}
      #1\quad & \mbox{if } #2 \\
      #3\quad & \mbox{if } #4 \\
      #5\quad & \IfValueTF{#6}{\mbox{if } #6}{\mbox{otherwise}}
    \end{array}
  \right.
}

\tikzset{robust/.style={line width=.16ex,line join=round}}

\newcommand*\rtlola{\text{RTLola}\xspace}
\newcommand*\lola{\text{Lola}\xspace}
\newcommand*\rust{\text{Rust}\xspace}
\newcommand*\viper{\text{Viper}\xspace}

\newcommand*\stl{\textsc{stl}\xspace}
\newcommand*\mtl{\textsc{mtl}\xspace}
\newcommand*\fpga{\textsc{fpga}\xspace}
\newcommand*\asic{\textsc{asic}\xspace}
\newcommand*\cpld{\textsc{cpld}\xspace}

\newcommand*\vhdl{\textsc{vhdl}\xspace}
\newcommand*\copilot{\text{Copilot}\xspace}

\newcommand*\preflen[1][\spec]{{\eta^{\leftarrow}_{#1}}} 
\newcommand*\postlen[1][\spec]{{\eta^{\rightarrow}_{#1}}} 

\makeatletter
\newcommand*\@evalorder{\leq_{\mathit{eo}}}
\newcommand*\evalorder{\mathrel{\@evalorder}}
\newcommand*\extevalorder{\mathrel{\@evalorder^+}}
\makeatother

\NewDocumentCommand{\shift}{ s g }{
  \IfNoValueTF{#2}{\Delta}{
    \IfBooleanTF{#1}%
    { \Delta\left({#2}\right) }%
    { \Delta({#2}) }%
  }
}

\NewDocumentCommand{\memreq}{ s g }{
  \IfNoValueTF{#2}{\mu}{
    \IfBooleanTF{#1}%
    { \mu\left({#2}\right) }%
    { \mu({#2}) }%
  }
}
\NewDocumentCommand{\Set}{ s m }{
  \IfBooleanTF{#1}%
  {\{#2\}}%
  {\left\{{#2}\right\}}%
}
\NewDocumentCommand{\card}{ s m }{
  \IfBooleanTF{#1}%
  {\vert#2\vert}%
  {\left\vert{#2}\right\vert}%
}
\newcommand*\given\mid

\definecolor{bluekeywords}{rgb}{0.13, 0.13, 1}
\definecolor{greentypes}{rgb}{0, 0.5, 0}
\definecolor{redstrings}{RGB}{171, 114, 2}
\definecolor{graynumbers}{rgb}{0.5, 0.5, 0.5}
\definecolor{goldcomments}{rgb}{0.6, 0.4, 0.08}
\definecolor{monitorblue}{RGB}{18, 163, 38}

\lstdefinelanguage{Lola}{
  keywords=[0]{input, output, trigger, constant},
  keywordstyle=[0]\bfseries\color{bluekeywords},
  keywords=[1]{if, then, else, aggregate, defaults, offset, hold},
  keywords=[2]{Int8, Int16, Int32, Int64, UInt8, UInt16, UInt32, UInt64, Bool, Float32, Float64, @1Hz, @5Hz, @10Hz, @100mHz, @1kHz},
  keywordstyle=[2]\color{greentypes},
    sensitive=false,
    comment=[l]{//},
    morecomment=[s]{/*}{*/},
    morestring=[b]',
    morestring=[b]"
}

\newcommand{\tikzmark}[1]{\tikz[overlay,remember picture] \node (#1) {};}
\lstdefinestyle{LolaDefault}{
    language={Lola},
    commentstyle=\color{goldcomments},
    keywordstyle=[1]\color{bluekeywords},
    stringstyle=\color{redstrings},
    numberstyle=\color{graynumbers},
    showstringspaces=false,
    basicstyle=\ttfamily\footnotesize,
}

\NewDocumentCommand{\bool}{O{}}{\mathds{B}^{#1}}

\usetikzlibrary{shapes,arrows,automata,backgrounds,patterns}

\tikzstyle{state} = [circle, draw, text width=.5cm, text centered, minimum height=.5cm, minimum width=.5cm]
\tikzstyle{component} = [rectangle, draw, text width=8em, text centered, minimum height=3em, fill=white]
\tikzstyle{smcomponent} = [component, minimum width=3cm, minimum height=2.5cm]
\tikzstyle{transition} = [draw, -stealth']
\tikzstyle{signal} = [draw, -latex']
\tikzstyle{config} = [densely dotted]

\colorlet{eventcolor}{green!50!black}
\colorlet{periodiccolor}{blue!50!black}
\tikzstyle{event} = [draw=eventcolor, thin, fill opacity=.3, pattern=north west lines, pattern color=eventcolor]
\tikzstyle{periodic} = [draw=periodiccolor, thin, fill opacity=.3, pattern=north east lines, pattern color=periodiccolor]
\tikzstyle{signalname} = [text width=10em, minimum height=2em]
\tikzstyle{nameright} = [signalname, align=left]
\tikzstyle{namecenter} = [minimum height=2em, align=center, rotate=65]
\tikzstyle{nameleft} = [signalname, align=right]

\newcommand*{\signal}[1]{\textsl{#1}}

\newcommand*{\component}[1]{\textsc{#1}}

\newcommand*{\sizets}{{s_{\ts}}}
\newcommand*{\sizeev}{{s_{\mathit{ev}}}}

\newcommand*{\numdl}{{\ensuremath{\#\mathit{dl}}}}

\newcommand\handcraftedclock[4][2]{%
  \begin{tikzpicture}[scale=#1,line cap=round,line width=#1*3pt]
    \filldraw [fill=periodiccolor!20!white] (0,0) circle (2cm);
    \foreach \angle / \label in
      {0/3, 30/2, 60/1, 90/12, 120/11, 150/10, 180/9,
      210/8, 240/7, 270/6, 300/5, 330/4}
    {
      \draw[line width=#1*1pt] (\angle:1.8cm) -- (\angle:2cm);
      \draw (\angle:1.4cm) node[scale=#1]{\textsf{\label}};
    }
    \foreach \angle in {0,90,180,270}
      \draw[line width=#1*2pt] (\angle:1.6cm) -- (\angle:2cm);
      \node[draw=none,font=\tiny,text=black,scale=#1] at (0,.9cm) {RV 2020};
      \draw[rotate=90,line width=#1*2pt] (0,0) -- (-#2*30-#3*30/60:0.7cm); 
      \draw[rotate=90,line width=#1*1.5pt] (0,0) -- (-#3*6:1cm); 
      \draw[rotate=90,line width=#1*.6pt,red] (0,0) -- (-#4*6:1.2cm); 
      \path [fill=black] (0,0) circle (3pt);
      \path [fill=red] (0,0) circle (1.5pt);
  \end{tikzpicture}%
}

\newcommand{\clockicon}[1]{\handcraftedclock[#1]{10}{8}{50}}

\definecolor{CommentColor}{rgb}{0.16,0.60,0.16}

\begin{document}

\title{Monitoring Cyber-Physical Systems: \\From Design to Integration\thanks{This work was partially supported by the German Research Foundation (DFG) as part of the Collaborative Research Center “Foundations of Perspicuous Software Systems” (TRR 248, 389792660), and by the European Research Council (ERC) Grant OSARES (No. 683300).}}

\author{Maximilian Schwenger
}
\institute{CISPA Helmholtz Center for Information Security \\
	\email{maximilian.schwenger@cispa.saarland}
}

\maketitle

\begin{abstract}
Cyber-physical systems are inherently safety-critical.
The deployment of a runtime monitor significantly increases confidence in their safety.
The effectiveness of the monitor can be maximized by considering it an integral component during its development.
Thus, in this paper, I given an overview over recent work regarding a development process for runtime monitors alongside a cyber-physical system.
This process includes the transformation of desirable safety properties into the formal specification language \rtlola.
A compiler then generates an executable artifact for monitoring the specification.
This artifact can then be integrated into the system.
\end{abstract}

\section{Introduction}
Cyber-physical systems (CPS) directly interact with the physical world, rendering them inherently safety-critical.
Integrating a runtime monitor into the CPS greatly increases confidence in its safety.
The monitor assesses the health status of the system based on available data sources such as sensors.
It detects a deterioration of the health and alerts the system such that it can \eg initiate mitigation procedures.
In this paper I will provide an overview regarding the development process of a runtime monitor for CPS based on recent work.
For this, I will use the \rtlola~\cite{rtlolaarxiv,rtlolacavtoolpaper} monitoring framework.

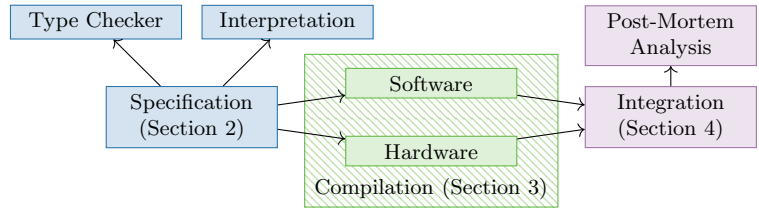
\begin{figure*}[t]
  \centering
\begin{tikzpicture}[yscale=-1,scale=.9, every node/.style={scale=.9}]
  
  \node [fill=loop!20, draw=loop, align=center, minimum width=2.5cm] (spec) at (0,0) {Specification\\ (\Cref{sec:specification})};
  \node [fill=loop!20, draw=loop, draw, align=center, minimum width=2.5cm] (tc) at (-1.4,-1.4) {Type Checker};
  \node [fill=loop!20, draw=loop, draw, align=center, minimum width=2.5cm] (inter) at (+1.4,-1.4) {Interpretation};
  \node [fill=prefix!20, draw=prefix, minimum width=3.7cm, minimum height=2.25cm, pattern=north west lines, pattern color=prefix!50] (comp) at (3.5, 0.2) {};
  \node [align=center] (compl) at (3.5,1.05) {Compilation (\Cref{sec:compilation})};
  \node [fill=prefix!20, draw=prefix, draw, align=center, minimum width=2.5cm] (hw) at (3.5,.5) {Hardware};
  \node [fill=prefix!20, draw=prefix, draw, align=center, minimum width=2.5cm] (sw) at (3.5,-.5) {Software};
  \node [fill=postfix!20, draw=postfix, draw, align=center, minimum width=2.5cm] (inte) at (7,0) {Integration\\ (\Cref{sec:integration})};
  \node [fill=postfix!20, draw=postfix, draw, align=center, minimum width=2.5cm] (pm) at (7,-1.2) {Post-Mortem\\ Analysis};
  
  \draw[->] (spec) edge (hw) (hw) edge (inte);
  \draw[->] (spec) edge (sw) (sw) edge (inte);
  \draw[->] (spec) edge (tc);
  \draw[->] (spec) edge (inter);
  \draw[->] (inte) edge (pm);
  
\end{tikzpicture}
  \vspace{-.5cm}
  \caption{Illustration of the paper structure. It is divided into three phases: specification, compilation, and integration.}
  \label{fig:paperstructure}
\end{figure*}

The process ranges from designing specifications to integrating the executable monitor.
It starts by identifying relevant properties and translating them into a formal specification language.
The resulting specification is type-checked and validated to increase confidence in its correctness.
Afterwards, it is compiled into an executable artifact, either based on software or hardware.
Lastly, the artifact is integrated into the full system.
This step takes the existing system architecture of the CPS into account and enables the monitor to support a post-mortem analysis.
The full process is illustrated in \Cref{fig:paperstructure}.


The first step of the process concerns the specification.
It captures a detailed analysis of the system behavior, which entails computationally challenging arithmetic.
Yet, since the monitor for the specification will be realized as an embedded component, its resource consumption must be statically bounded.
Thus, the specification language has to provide sufficient expressiveness while allowing the monitor to retain a predictable and low resource footprint. 
In particular, an ideal specification language provides formal guarantees on the runtime behavior of its monitors such as worst case execution time or memory consumption.
In general, however, expressiveness, formal guarantees, and predictably low resource consumption cannot be achieved at the same time.
Desirable properties like ``every request must be granted within a second'' might come at the cost that the memory consumption of the monitor depends on the unpredictable input frequency of requests.
Consequently, specification languages providing input-independent formal guarantees on the required memory must impose restrictions to prohibit such properties.
These restriction can be direct, \ie, the syntax of the language renders the property inexpressible, or indirect, so the property can be expressed but falls into a language fragment unsuitable for monitoring. 
\rtlola falls into the former category.

During the design phase of the CPS, the specifier defines properties spanning from validation of low-level input sensor readings to high-level mission-specific control decisions. 
The former are real-time critical, \ie, they demand a timely response from the monitor, whereas the latter include long-term checks and statistical analysis~\cite{rtlolacavindustrial} where slight delays and mild inaccuracies are unsubstantial.
Just like code is not a perfect reflection of what the programmer had in mind, the specification might deviate from the specifiers intention.
To reduce the amount of undetected bugs, the specification needs to be validated. 
This increases confidence in it and --- by proxy --- in the monitor. 
The validation consists of two parts:  type checks and validation based on log data.
The former relies solely on the specification itself and checks for type errors or undefined behavior.
The latter requires access to recorded or simulated traces of the system and interprets the specification over the given traces.  
The output of the monitor can then be compared against the expected result.

After successfully validating the specification, a compiler for the specification language generates an executable artifact.
This artifact is either a hardware or a software solution, depending on the requirements of the system architecture. 
If, for example, the architecture does not allow for adding additional components, a software solution is preferable as it does not require dedicated hardware; the monitor can be part of the control computer.  
Hardware solutions, on the other hand, are more resource efficient and allow for parallelization with nearly-0 cost.
In any case, the compiler can inject additional annotations for static code-level verification\cite{lolatorust} or traceability\cite{janmaster} to further increase confidence in the correctness of the monitor.

Finally, deploying the monitor into the CPS harbors additional pitfalls. 
As an external safety component, the monitor should not influence the regular operation of the system unless upon detection of a safety violation.
As a result, the system architecture needs to enable non-intrusive data flow from the system to the monitor and intrusive data flow from the monitor to the controller.
The controller then has to react on an alarm appropriately. 
Such a reaction can \eg be a switch from the regular controller to a formally verified controller with significantly reduced complexity responsible for a graceful shutdown of the system~\cite{mtcps20}, as suggested in the Simplex Architecture~\cite{simplex}. 

After terminating a mission, the output of the monitor provides valuable data for the post-mortem analysis. 
Regular system logs might be insufficient as they do not contain the entire periphery data due to resource constraints. 
The monitor, however, filters and aggregates the data specifically to assess information regarding the system's status \wrt safety, thus providing valuable feedback.

\section{Specifications: From Sensors to Missions}\label{sec:specification}

When designing specifications for CPS, the specifier has to keep in mind that not all properties are equal.
They fall into a spectrum from low-level properties concerned with concrete sensor readings to high-level properties validating mission-specific criteria.
Properties on the least end of the spectrum work on raw data points of single sensors.  
Most common are simple bounds checks (\textit{the altitude may not be negative}) or frequency checks (\textit{the barometer must provide between 9 and 11 readings per second}).
Less low-level properties work on refined data points, \eg to check whether several sensors contradict each other (\textit{the altitude measured by the sonic altimeter must not deviate more than $\epsilon$ from the altitude based on the air pressure}).
Such a sensor cross-va\-li\-da\-tion requires refinement of raw values as they cannot be compared without further ado.
While a barometer provides the air pressure, it needs further information such as pressure and temperature at sea level to accurately estimate the current altitude.
Similarly, validating the position provided by the \emph{global navigation satellite system} (GNSS) module against the position estimated by the \emph{inertial measurement unit} (IMU) requires double integration of the measured acceleration.
On the highest end of the spectrum reside mission-level properties.
When checking such properties, the source of information is mostly discarded and the values are assumed to be correct.
For example, consider an aircraft that traverses a set of dynamically received waypoints.
Mission-level properties could demand that a waypoint is reached in time or that the traveled distance does deviate more than a factor from the actual distance between two points.

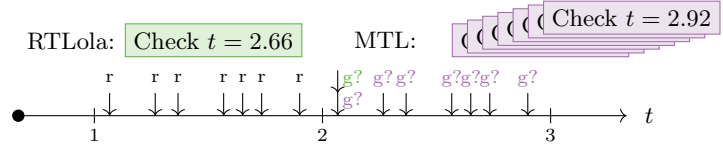
\begin{figure*}[t]
  \centering
\begin{tikzpicture}[yscale=1,scale=1, every node/.style={scale=1}]

  \draw[] (0,0) edge[->] (8,0);
  \node[] (t) at (8.3, 0) {$t$};
  \node[draw, fill, circle, inner sep=1.5pt] (dot) at (0,0) {};
  
  \foreach \x in {1, 2, 3} {
    \draw[] (\x*3-2, .1) -- ++(0, -.2);
    \node[] (boop) at (\x*3-2, -.25) {\scriptsize \x};    
  }
  
  \foreach \x in {1.2, 1.8, 2.1, 2.7, 2.95, 3.2, 3.7} {
    \draw (\x, .3) edge[->] (\x, 0);
    \node[] (req) at (\x, .5) {\scriptsize r};
    \draw[draw=postfix] (\x+3, .3) edge[->] (\x+3, 0);    
  }
  \foreach \x in {1.8, 2.1, 2.7, 2.95, 3.2, 3.7} {
    \node[] (grant) at (\x+3, .5) {\scriptsize\color{postfix} g?};
  }
  \node[] (grant) at (1.2+3+.2, .2) {\scriptsize\color{postfix} g?};
  \node[] (grant) at (1.2+3+.2, .5) {\scriptsize\color{prefix} g?};
  \draw[draw=prefix] (4.2, .6) edge[->] (4.2, 0.3);
  
  \node[anchor=west] (rtlola) at (0, 1) {\rtlola:};
  \node[fill=prefix!20, draw=prefix, anchor=west] (rtlolacheck) at (1.4, 1) {Check $t=2.66$};
  \node[anchor=west] (rtlola) at (4.3, 1) {MTL:};
  \foreach \x in {0,1,2,3,4,5,6} {
    \node[fill=postfix!20, draw=postfix, anchor=west] (mtlcheck) at (5.7+\x*.2, 1+\x*.05) {Check $t=2.92$};
  }
%
%
  
\end{tikzpicture}
  \vspace{-0.5cm}
  \caption{Illustration of monitor obligations for checking if a request (``r'') is granted (``g?'') within a second. While the \mtl interpretation is more precise, it requires statically unbounded memory.  The \rtlola under-approximation requires constant memory.}
  \label{fig:costofexpressiveness}
\end{figure*}

Properties are usually expressed in natural language as above and translated into a specification language.
Consider the first property: \textit{the altitude may not be negative}.
Evidently, the properties harbors little challenge in terms of arithmetic.
However, timeliness is critical. 
If the altimeter reports a negative altitude, clearly something is wrong and the system needs to be informed near-instantaneously.
In \rtlola, the property translates to the following specification:
\begin{lstlisting}
  input altitude: Float32
  input orientation: Float32
  trigger altitude < 0 "Altimeter reports negative altitude."
  trigger orientation > 2 * $\pi$ "Orientation exceeds 2$\color{redstrings}\pi$."
\end{lstlisting}
The first two lines declare input streams of name \lstinline{altitude} and \lstinline{orientation}, both with type \lstinline{Float32}.
The remaining lines contain trigger conditions with message to be sent to the system for the case a condition turns true.
Whenever the monitor receives a new value from the altimeter or gyroscope, the respective condition is checked immediately.
Note that \rtlola allows for \emph{asynchronous} input behavior, \ie, one input stream can produce a new value while the other does not.
Thus, when the gyroscope produces a value, the respective trigger condition is checked regardless of the altimeter.
This timing dependency from inputs to expressions is part of \rtlola's type system.

The type system is two-dimensional:  every stream and expression has a \emph{value type} and a \emph{pacing type}.
Value types are common within programming languages, they indicate the shape and interpretation of data.
The input streams, for example, are of value type \lstinline{Float32}, so storing a single value requires 32 bits and the bits should be interpreted as a floating point number.
The pacing type, however, states \emph{when} expressions are evaluated and thus when streams produce new values.
In case of the trigger expressions, the pacing types are \emph{event-based}, \ie, they are coupled to the reception of new values from the altimeter or gyroscope.

The pacing type can also be \emph{periodic}, effectively decoupling the evaluation of expressions from input streams in terms of timing.
As an example, consider the second low-level property: \textit{the barometer must provide between 9 and 11 readings per second}.
An \rtlola specification for this property is:
\begin{lstlisting}
input pressure: Float32  
output readings_per_sec @ 1Hz := pressure.aggregate(over: 1s, using: count)
trigger readings_per_sec < 9 "Barometer produces too few readings."
trigger readings_per_sec > 11 "Barometer produces too many readings."
\end{lstlisting}
Here, \lstinline{readings_per_sec} is an output stream with a timing annotation \lstinline{@ 1Hz} prompting the monitor to only evaluate the expression of the stream once per second.
Thus, the timing of the evaluation is decoupled from the reception of new input values.

The expression itself is a sliding window aggregation, \ie, whenever evaluated, the expression counts how many data points the barometer generated in the last second.
If the count falls below 9 or exceeds 11, the monitor raises an alarm.
While the logic behind a an efficient sliding window implementation is rather complex and requires a great deal of bookkeeping, \rtlola provides a simple primitive for it.
This alleviates the need for the specifier to manually take care of the implementation details.

Note that the specification does not precisely represent the property.
Assume the system alternates between receiving 7 readings in the first half of a second and receiving 3 readings in the second half.
Then, every other second, the system receives a total of 14 readings per second --- unbeknownst to the monitor.
In \rtlola, it is impossible to specify the property correctly as it lacks the necessary expressiveness by design: sliding window expressions can only occur in streams with a timing annotation.
This annotation renders the stream \emph{isochronous}, \ie, the point in time when its expression will be evaluated are determined statically.
The reason behind it is that the original properties lies in a category of properties that generally need a statically unbounded amount of memory to be monitored.
To understand this, consider the property \textit{Every request must be granted within one second}.
A sound monitor for the property needs to check whether a request was granted exactly one second after receiving a request.
However, there is no static bound on the amount of requests the monitor receives within this frame of time.
Since it has to store their arrival times, the memory consumption might exceed any bound.
The problem is illustrated in \Cref{fig:costofexpressiveness}.
There are specification logics such as \emph{metric temporal logic} (\mtl)~\cite{mtl}, in which the property can be expressed.
In such a case, the memory consumption of the monitor is linear in the number of events receives within the second.
Since \rtlola only provides constant memory monitors, it rejects specifications for such properties and instead enables constant-memory under-approximations.
This design decision is a requirement to guarantee that the monitor cannot possible run out of memory during the execution.

\rtlola provides primitives for more abstract constraints such as sensor cross-validations as well:
\begin{lstlisting}
  input velocity_1: Int64
  input velocity_2: Int64
  output deviation := abs(velocity_1 - velocity_2)
  output lasting_dev := deviation > 5 $\land$ deviation.offset(by: -1, default: 0) > 5 $\land$ deviation.offset(by: -2, default: 0) > 5 
  trigger lasting_deviation "Lasting deviation in measured velocities."
\end{lstlisting}
The specification declares two input streams providing different readings for the velocity of the system, and two output streams \lstinline{deviation} and \lstinline{lasting_dev} that computes the absolut deviation of the readings and checks whether the deviation exceeds a threshold three consecutive times.
The first conjunct of the stream expression accesses the current, \ie, the latest value of the \lstinline{deviation} stream, whereas the \lstinline{offset(by: -n, default: v)} function allows for accessing the $n$th-to-latest value of the stream for $n \in \mathds{N}$\footnote{As a result, \rtlola does not allow for accessing future values.}.
This value does not exist at the beginning of the monitor execution, so the specifier has to supply a default value~$v$. 
Here, the specification refers to the abstract notion of ``the last value'' rather than considering the real-time behavior, assuming that low-level validation already took place.

Note that \lstinline{deviation} accesses both \lstinline{velocity} streams without supplying a default value.
This indicates a \emph{synchronous} access and prompts the monitor to only evaluate \lstinline{deviation} when both input receive a new value.
This is not necessarily the case since \rtlola considers inputs to be asynchronous.
The pacing type of \lstinline{deviation} captures the information that the stream is only evaluated when the two inputs happen to arrive at the same time:  it is event-based and the conjunction of both input streams.
In contrast to that, a different definition of \lstinline{deviation} could look like:

\begin{lstlisting}
output deviation_disj @ velocity_1 $\lor$ velocity_2 := abs(velocity_1.hold(or: 0) - velocity_2.hold(or: 0)  
\end{lstlisting}

Here, the output stream has a disjunctive type, so when it is extended, at least one of the two inputs received a new value, not necessarily both.
In such a case, \rtlola forces the specifier to declare precisely how it should handle potentially old values.
The specifier can, as in the example of \lstinline{deviation_disj}, turn the synchronous accesses into sample and hold accesses.
When evaluating the expression, the monitor will access the latest --- yet potentially old --- value of the input stream with a 0-order hold.
If the specifier attempted to access either stream synchronously, \rtlola would reject the specification because it contains an inner contradiction.
These kinds of type checks greatly increase confidence in the correctness of the specification as they point out imprecise and potentially flawed parts.

Lastly, consider two mission-level properties for an aircraft flying a dynamic list of waypoints: the monitor should issue a warning if the aircraft deviated from the straight-line distance by at least $\varepsilon$, and it should issue an error if such a deviation occurred more than 10 times.

\begin{lstlisting}
input wp: (Float64, Float64)
input pos: (Float64, Float64)
output start: (Float64, Float64) := (0, 0)
output exp_dist @ wp := wp - wp.offset(by: -1, default: start)
output dist_since_wp @ pos := pos - pos.offset(by: -1, default: start)
  + dist_since_wp.offset(by: -1, default: 0)
output distance_deviation @ wp := 
  abs(exp_dist.offset(by: -1, default: 0) - dist_since_wp.hold(or: 0))
trigger distance_deviation > $\varepsilon$ "Warn: Path deviation detected."
output deviations := deviations.offset(by: -1, default: 0) 
  + if distance_deviation > $\varepsilon$ then 1 else 0
trigger deviations > 10 "Err: Too many path deviations!"
\end{lstlisting}

The specification declares two clearly asynchronous input streams: the current waypoint \lstinline{wp} and the current position \lstinline{pos}.
The \lstinline{start} stream is a constant stream only containing the initial position of the aircraft.
\lstinline{exp_dist} contains the distance between the current and the last waypoint whereas \lstinline{dist_since_wp} aggregates the distance the aircraft has traveled since reaching the last waypoint/starting the mission.
The deviation in distance is then the absolut difference between these two distances.
Note that this value is only valid when the aircraft just reached a new waypoint, hence the \lstinline{@ wp} annotation.
This prompts the monitor to only evaluate the stream when it receives a new waypoint.
Lastly, the \lstinline{deviations} stream counts the number of times the deviation exceeded its threshold.

The specification contains several pacing type annotations.
This, however, is for illustration, as most of the time, \rtlola can infer both types from the stream expression.
Yet, the specifier always has the option to annotate types for clarity or if the timing of a stream should deviate from the standard behavior, \eg for disjunctive event-based types.

Note that this was only a brief overview of \rtlola.  
For more details on the theory, refer to~\cite{maxmaster}, and for the implementation, refer to~\cite{rtlolacavtoolpaper}.

\subsection{Specification Validation by Interpretation}

\rtlola's type system already rules out several sources for incorrect behavior of the monitor.
Yet, a validation of the specification is crucial to increase confidence in the \emph{correctness} of the specification.
The validation requires access to records of previous runs of the system.
These can be simulated, collected during test runs, or logs from sufficiently similar systems.
Just like when testing software, developers annotate the trace data with points in time when they expect the monitor to raise an alarm.
Then, they execute a monitor for the given specification on the trace and compare the result with their annotations.
Deviations mainly root from either an error when designing the specification, or a discrepancy in the mental image of different people regarding the correct interpretation of a property.

A key point for the specification validation is that the process should incur as little cost as possible to enable rapid prototyping.
Hence, interpreting the specification rather than compiling it is preferable --- especially when the target platform is hardware-based.
After all, realizing a specification on an \fpga usually takes upwards of \SI{30}{\minute}~\cite{fpgalola}.
While interpretation is considerably less performant than compiled solutions, the \rtlola interpreter manages to process a single event in \SI{1.5}{\micro\second}.
This enables a reasonably fast validation of specifications even against large traces.

\subsection{Static Analysis for RTLola Specifications}\label{sec:static:analyses}

After type checking the specification and validating its correctness based on test traces, \rtlola provides static checks to further analyze it.
For this, \rtlola generates a dependency graph where each stream is a node and each stream access is an edge.
This information suffices to perform a memory analysis and a running time analysis.
The analysis identifies the resource consumption --- both spatial and temporal --- of each stream, granting fine-grained control to the specifier.

\paragraph{Memory}
For the memory consumption of stream $s$, the analysis identifies the access of another stream to $s$ with the greatest offset $n^\ast_s$. 
Evidently, the monitor only has to retain $n^\ast_s$ values of $s$ to successfully resolve all accesses.
Moreover, note that all types in \rtlola have a fixed size.  
Let $T_s$ be the type of $s$ with bit-size $\card{T_s}$.
Then, the memory consumption of $s$ induced by accesses through other streams amounts to $n^\ast_s \cdot \card{T_s}$.

Sliding window expressions within the stream expression of $s$ incur additional memory overhead. 
Suppose $w_1,\dots,w_k$ are the windows occurring in $s$ where for $w_i = (\gamma_i, d_i)$, $\gamma_i$ is the aggregation function and $d_i$ is the length of the window.
If $k > 0$, \rtlola demands $s$ to be periodic with frequency $\pi_s$.
The memory consumption of $s$ induced by sliding windows consists of the number of \emph{panes} required.
Here, a pane represents the time interval between two consecutive evaluations of the window.
The pane consists of a single value which contains the aggregated information of all values that arrived in the respective time interval.
This implementation of sliding windows is inspired by Li~\etal~\cite{nopane} and only works for list homomorphisms~\cite{meertens}.
A window thus has $d_i\cdot\pi_s$ panes, which has to be multiplied by the size of the value stored within a pane: $T_{\gamma_i}$.
This value is statically determined and depends on the aggregation function:  for summation, it is merely the sum of the values, for the average it is the intermediate average plus the number of values that occurred within the pane.

The overall memory consumption of $s$ is therefore
\[ \mu(s) = n^\ast_s \card{T_s} + \mathds{1}_{k>0} \sum_{i=1}^k d_i \pi_i \card{T_{\gamma_i}} \]
Here, $\mathds{1}_\varphi$ is the indicator function evaluating to 1 if $\varphi$ is true and 0 otherwise.

\paragraph{Running Time}
The running time cannot be fully determined based on the specification alone as it depends on the hardware of the CPS.
For this reason, \rtlola provides a preliminary analysis that can be concretized given the concrete target platform.
The preliminary analysis computes 
\begin{enumerate*}[label=\alph*)]
\item the complexity of each evaluation cycle given a certain event or point in time, and
\item the parallelizability of the specification.
\end{enumerate*}

For the former metric, note that the monitor starts evaluation cycles either when it receives an event, or at predetermined points in time (\emph{deadlines}).
An event always updates a set of input streams and a statically determined set of output streams.
Recall the mission-level specification computing the deviation from the flight path including the \lstinline{deviation_disj} stream.
The specification declares two input streams, thus allowing for three possible non-empty events.
An event covering either \lstinline{velocity} stream but not the other only triggers an evaluation of \lstinline{deviation_disj}.  
Only if the event covers both inputs, \lstinline{deviation} and the trigger are evaluated as well.

Consider a specification containing periodic streams and suppose the monitor has a deadline at time $t$.
It then evaluates all periodic streams due at $t$, \ie, all streams with frequency $\pi$ where $\pi \cdot t$ is a natural number.
Thus, the set of streams affected by an evaluation cycle is pre-determined.  

The next step in the analysis is concerned with \emph{evaluation layers}
They are closely related to the parallelizability of the monitor as they indicate how many stream evaluations can take place at once.
The analysis yields a partition of the set of streams where all streams within an element of the partition are \emph{independent}, enabling a parallel evaluation.
The (in-)dependence relation is based on the dependency graph.
If a stream accesses another \emph{synchronously}, \ie, without an offset, than the target stream needs to be evaluated before the accessing stream.
This definition entails an \emph{evaluation order} on output streams.
The aforementioned partition is then the coarsest partition such that any two streams in the same set are incomparable with respect to the transitive closure of the evaluation order.
Each element of the partition is an evaluation layer.
By construction, streams within the same layer can be evaluated in an arbitrary order --- in particular also in parallel.
The order in which layers are evaluated, however, still needs to follow the evaluation order.
In the example specification before, the partition would be \lstinline[language=]!{{wp, pos, start} < {exp_dist, dist_since_wp} < {distance_deviation} < {deviations, trigger_warn} < {trigger_err}}!.  

The evaluation layer analysis immediately provides information regarding the parallelizability of the monitor.
The running time analysis takes the number of evaluations into account as well as how many streams are affected by an evaluation cycle, and how complex their expressions are.
Intuitively, if an event or deadline affects streams in a multitude of layers, then the evaluation is slow as computations depend on each other and thus require a sequential order.
Conversely, if an event only affects few streams, all within the first layer, the evaluations are independent and thus highly parallelizable.
As a result, the running time of the monitor is low.

Note, however, that for software monitors the degree to which computations should run in parallel requires careful consideration, since spawning threads incurs a constant overhead.
For hardware monitors, the overhead does not apply.

\section{Compilation: Generating Monitors}\label{sec:compilation}

While interpretation of specifications enables rapid prototyping, its logic is far more complex than a compiled monitor, at the same time resulting in subpar performance.
This renders compilation preferable.
Additionally, the compiler can inject additional information into the generated code.
Such annotations can benefit the certification process of the CPS either by providing a notion of traceability, or by outright enabling the static verification of the monitor.
The target platform of the compilation can either be hardware or software, both coming with advantages and drawbacks.

In this section, I will present and discuss a hardware compilation for \rtlola specifications, and a software compilation for \lola, \ie, a subset of \rtlola, with verification annotations.

\subsection{RTLola on FPGA}

Realizing an \rtlola specification on hardware has several advantages.
For one, the hardware monitor does not share resources with the controller of the system apart from power, eliminating potential negative interference.
Moreover, special purpose hardware tends to be smaller, lighter, and require less energy than their general purpose counterparts.
Secondly, hardware enables parallel computations at minimal cost.
This synergizes well with \rtlola, where output streams within the same evaluation layer can be evaluated in parallel.

The realization of \rtlola on hardware~\cite{fpgalola} works in two steps: an \rtlola specification is translated into \vhdl code, out of which an \fpga (field-pro\-gram\-ma\-ble gate array) implementation is synthesized.
The synthesis provides additional static information regarding the required board size in terms of memory\footnote{The hardware realization might require temporary registers and working memory. This can slightly increase the computed memory consumption.} and lookup-up tables.
This allows for validating whether the available hardware suffices to host the monitor.
Moreover, the synthesis indicates the idle and peak power consumption of the monitor, information that is invaluable when integrating the monitor into the system.

The aforementioned advantages are not only valid for \fpga but also for other hardware realization such as application-specific integrated circuits (\asic) and complex programmable logic device (\cpld).
While \asic have significant advantages over \fpga when it comes to mass-producibility, power consumption and performance, \fpga are preferable during the development phase as they are orders of magnitude cheaper, have a lower entry barrier, and allow for rapid development.
\cpld, on the other hand, are just too small to host realistic/non-trivial specifications.

\subsubsection{Hardware Realization}

Managing periodic and event-based streams under a common hardware clock poses the key challenging when realizing an \rtlola monitor in hardware.
Yet, this distinction only affects the part of the monitor logic deciding \emph{when} to evaluate stream expressions; the evaluation itself is agnostic to it.
For this reason, the monitor is split into two modules.
The \emph{high-level controller} (\hlc) is responsible for scheduling evaluations, \ie, to decide when and which streams to evaluation.
It passes the information down to the second module, the \emph{low-level controller} (\llc), which is responsible for managing the evaluation.
A \textsc{fifo} queue between the controllers buffers information sent from the \hlc to the \llc.

Recall the specification from \Cref{sec:specification} checking for strong deviations in readings of two velocimeters.
As a running example for the hardware compilation, we extend the specification by the following declarations.
\begin{lstlisting}
  output avg_dev @10mHz := dev.aggregate(over: 10min, using: avg)
  trigger avg_dev > 4 "High average deviation."
\end{lstlisting}
The specification checks whether two velocimeters disagree strongly over three consecutive measurements and whether the average disagreement is close to the disagreement-threshold.
Note that all streams are event-based except for \lstinline{avg_dev} and the second trigger.

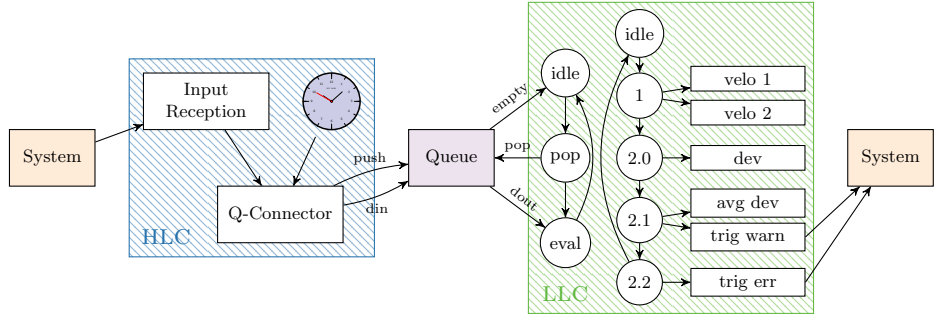
\begin{figure}[t]
  \tikzstyle{wecomp} = [draw, align=center, minimum height=1cm, minimum width=1.5cm, fill=white]
\tikzstyle{westate} = [draw, circle, minimum width=.8cm, fill=white]
\tikzstyle{westream} = [draw, minimum width=2cm, fill=white]

\centering
\begin{tikzpicture}[yscale=-1,scale=.75, every node/.style={scale=.75}, >=stealth']

  \node [wecomp, fill=prelude!20] (system1) at (0,0) {System};
  
  \node [draw=loop, minimum width=4.3cm, minimum height=3.5cm, pattern=north west lines, pattern color=loop!50] (hlc) at (3.5, 0) {};
  \node [wecomp, minimum width=2.2cm] (inp) at (2.7, -1) {Input\\ Reception};
  \node [] at (4.9, -1) (time) {\clockicon{.25}};
  \node [wecomp, minimum width=2.2cm] (qcon) at (4, 1) {Q-Connector};
  \node [] (hlclable) at (2, 1.4) {\Large\color{loop} \hlc};
  
  \node [wecomp, fill=postfix!20] (q) at (7,0) {Queue};
  
  \node [draw=prefix, minimum width=5cm, minimum height=5.5cm,pattern=north west lines, pattern color=prefix!50] (llc) at (10.85, 0) {};
  \node [] (llclable) at (9, 2.4) {\Large\color{prefix} \llc};

  \node [westate] (idle1) at (9, -1.5) {idle};
  \node [westate] (pop) at (9, 0) {pop};
  \node [westate] (eval) at (9, +1.5) {eval};

  \node [westate] (idle2) at (10.3, -2.2) {idle};  
  \node [westate] (1) at (10.3, -1.1) {1};
  \node [westate] (20) at (10.3, 0) {2.0};
  \node [westate] (21) at (10.3, +1.1) {2.1};
  \node [westate] (22) at (10.3, +2.2) {2.2};

  \node [westream] (velo1) at (12.2, -1.4) {velo 1};  
  \node [westream] (velo2) at (12.2, -0.8) {velo 2};  
  \node [westream] (dev) at (12.2, 0) {dev};  
  \node [westream] (avg) at (12.2, 0.8) {avg dev};  
  \node [westream] (trig1) at (12.2, 1.4) {trig warn};  
  \node [westream] (trig2) at (12.2, 2.2) {trig err};
  
  \node [wecomp, fill=prelude!20] (system2) at (14.7,0) {System};
  
  \path[->]
    (system1) edge (inp)
    (inp) edge (qcon)
    (time) edge (qcon)
    (qcon) edge[bend right=10] node[above] {\scriptsize push} (q)
    (qcon) edge[bend left=10] node[below] {\scriptsize din} (q)
    (q) edge node[align=left, sloped, above, rotate=75] {\scriptsize empty} (idle1)
    (pop) edge node[above] {\scriptsize pop} (q)
    (q) edge node[align=left, sloped, above, rotate=-75] {\scriptsize dout} (eval)
    (trig1.east) edge (system2)
    (trig2.east) edge (system2)
  ;
  \path[->]
    (idle1) edge (pop)
    (pop) edge (eval)
    (eval) edge[bend left=26] (idle1)
    (idle2) edge (1)
    (1) edge (20)
    (20) edge (21)
    (21) edge (22)
    (22) edge[bend right=27] (idle2)
    (22) edge (trig2)
    (21) edge (trig1)
    (21) edge (avg)
    (20) edge (dev)
    (1) edge (velo1)
    (1) edge (velo2)
    ;

%
  
\end{tikzpicture}
  \caption{Illustration of the hardware realization of an \rtlola monitor. It is composed of two components connected by a queue. 
  The \hlc receives inputs from the system and manages the timing of periodic and event-based streams. The \llc controls the evaluation process with a state machine where each state represents an evaluation layer of the underlying specification. The \llc passes violations of safety properties on to the system.}
  \label{fig:schematic:overall}
\end{figure}

\Cref{fig:schematic:overall} illustrates the overall structure of the monitor.
As can be seen, the \hlc accepts input events from the monitored system.
Such an event has a fixed size of $2\cdot(32+1)$ bits, \ie, 32 due to the input types and an additional bit per stream to indicate whether the stream received an update.
For periodic streams, the \hlc has access to the system clock.
Based on the current time and arrival of events, the \hlc triggers evaluation cycles by sending relevant information to the queue while rising the \signal{push} signal.
Such a data package consists of $2\cdot(32+1) + 64 + 5$ bits. 
The first component of the sum represents the potential input event.
If the evaluation only serves to update periodic streams, these bits will all be 0.
The following 64 bits contain the timestamp of the evaluation, crucial information for the computation of sliding window expressions.
The last 5 bits each represent an output stream or trigger and indicate whether the respective entity is affected by the evaluation cycle.
As a result, the \llc does not have to distinguish between event-based and periodic streams; it merely has to evaluate all streams the \hlc marked as affected.

The communication between the queue and the \llc consists of three data lines:
the \signal{pop} bit is set by the \llc and triggers the queue to send another data packet down to it --- provided the \signal{empty} bit is 0.  
In this case, the queue puts the oldest evaluation information on the \signal{dout} wires.

Internally, the \llc consists of two state machines.  
The first one handles the communication with the queue.
While the first machines resides in the \textit{eval} state, the second state machine manages the evaluation.
To this end, it cycles through different states, each representing an evaluation layer.
The first state (``1'') copies the information about input stream updates into the respective memory region.
In each consecutive state, the monitor enables the modules responsible for evaluating the respective stream expression by raising the \signal{enable} bit.
It then waits on the \signal{done} bits.
Upon receiving all of them, the monitor proceeds to the next state.
During this process, the outputs of trigger expressions are not persisted locally, but directly piped down to the monitored system.

\subsubsection{Resource Consumption}

When compiling the specification into \vhdl and realizing it on a \textsc{zynq-7 zc702} Evaluation Board, using the Vivado Design Suite\footnote{https://www.xilinx.com/products/design-tools/vivado.html}, the hardware synthesizer provides information regarding the overall resource consumption.
In this case, the monitor requires 10,700 lookup tables and 1735 bits of memory.
The energy consumption amounts to \SI{144}{\micro\watt} when idle, and \SI{1.699}{\watt} under peak pressure.
Even though the specification is rather small, it gives a glimpse at how low the resource consumption actually is.
Baumeister~\etal~\cite{rtlolacavindustrial} successfully synthesized larger specifications designed for autonomous aircraft on the same \fpga.

\subsection{Lola to Rust}


While a compilation of \rtlola specifications into software is a topic for future work, a compilation for \lola does exist, presented by Finkbeiner~\etal~\cite{lolatorust}.
\lola~\cite{lola} is a synchronous and discrete-time subset of \rtlola.
As such, it does not have a notion of real-time, thus neither sliding windows nor periodic streams are an issue.
Moreover, \lola assumes all input streams to receive new values at the same time, prompting all output streams to be extended as well.
This renders sample and hold accesses obsolete.
\lola does, however, allow for future lookups, \ie, a stream may refer to the \emph{next} value of another stream.

The example specification for the software compilation is another mild variation of the velocimeter cross-validation from \Cref{sec:specification}.
The modification replaces the \lstinline{lasting_dev} stream by the following:
\begin{lstlisting}
output lasting_dev := dev > 5 $\land$ dev.offset(by: +1, default: 0) > 5 $\land$ dev.offset(by: -1, default: 0) > 5 
\end{lstlisting}
Here, \lstinline{lasting_dev} access the last, current and \emph{next} value of \lstinline{deviation}.

The compilation presented in this section translates the \lola specification into Rust\footnote{https://www.rust-lang.org/} code that enables a static verification.
Rust as a target language comes with several advantages.
First, as a system language with an \textsc{LLVM}\footnote{https://llvm.org/} backend, it is compatible with a wide array of platforms.
Secondly, a key paradigm of the language is to enforce static checks on the code and thus reduce dynamic failures. 
This goes hand in hand with the goal of verifying the functional correctness and absence of dynamic failures of the generated monitor.
Lastly, Rust allows for fine-grained control low-level constructs such as memory management, enabling the programmer --- or in the case, the \lola compiler --- to write highly performant code.

The compiler injects verification annotations into the code as well.
This enables the static verifier \viper to prove functional correctness of the monitor in two steps. 
First, it relates the working memory of the monitor to the semantic model of \lola.  
The key challenge here is that the semantics of \lola argues about infinite data sequences while the monitor operates on a finite working memory.
Next, the verification relates the verdict of the monitor, \ie, the boolean indicator of whether a trigger should go off is correct given the current state of the working memory.
In combination we can conclude that the monitor only emits an alarm if the semantic model demands so.

\subsubsection{Dealing with Fallible Accesses}

While future offsets provide a natural way to specify temporal dependencies, the monitor has to compensate for them by delaying the evaluation of the accessing streams. 
Thus, the evaluation of \lstinline{lasting_dev} needs to be delayed by one step since they access a future value of \lstinline{dev}.
This delay is propagated through the dependency graph: the trigger transitively accesses a future value, so its evaluation needs to be delayed, too.

With the delay operation in place, accesses via a future offset will always succeed up until the system terminates, and thus no longer produces new inputs.
In this case, the monitor continues to evaluate delayed streams until they have the same length as the input streams.
This phase is the \emph{postfix} phase of the monitor execution. 
Here, future offsets fail because the accesses values do not exist and never will.
Similarly, past offsets fail at the beginning of the monitor execution, the \emph{prefix}.

In the very first iteration of the monitor, only the inputs and \lstinline{dev} can be evaluated, the other output stream and the trigger are delayed.
In the next iteration, the input is updated and all output streams and the trigger are evaluated.
Evaluating \lstinline{lasting_dev} accesses both values of \lstinline{dev}.
In addition to that, the past lookup refers to the -1st value of \lstinline{altitude}, a value, that will never exist.
Thus, the monitor statically substituted the access with the default value.

Clearly, the monitor goes through three phases: a prefix, in which past offsets fail unconditionally, a loop phase, in which both past and future offsets succeed unconditionally, and a postfix phase, in which future offsets fail unconditionally.
In light of this, the compiler faces a trade-off:  it can generate a general-purpose loop containing conditional statements resolving offsets dynamically, or it can take the three phases into account by generating code specific to them.
The former option contains conditional statements not found in the original specification, resulting in far less performant code.
The latter option, however, requires more code, resulting in a larger executable file.
The compiler outlined in this section opts for the latter option.

\begin{figure}[t] 
  \begin{tikzpicture}[]
	 \draw[rounded corners, fill=prelude!20, draw=prelude] (-2.1, 0) rectangle ++(6, -3.75);
	 \node[anchor=north east, inner sep=0, outer sep=0] (prelude) at (3.7, -.1) {\scriptsize\color{prelude}Prelude};
	 \draw[rounded corners, fill=loop!20, draw=loop] (-2.1, -4.3) rectangle ++(6, -1.15);
	 \node[anchor=south east, inner sep=0, outer sep=0] (loop) at (3.7, -5.35) {\scriptsize\color{loop}Monitor Loop};
	 \draw[rounded corners, fill=prefix!20, draw=prefix] (4.3, 0) rectangle ++(5.8, -3.3);
	 \node[anchor=south east, inner sep=0, outer sep=0] (pre) at (9.9, -3.2) {\scriptsize\color{prefix} Execution Prefix};
	 \draw[rounded corners, fill=postfix!20, draw=postfix] (4.3, -3.35) rectangle ++(5.8, -1.5);	 
	 \node[anchor=south east, inner sep=0, outer sep=0] (post) at (9.9, -4.75) {\scriptsize\color{postfix}Execution Postfix};
\end{tikzpicture}

\vspace{-5.75cm}
\begin{lstlisting}[,style=ColoredRust, basicstyle=\scriptsize,multicols=2, frame=none]
struct Memory { ... } $\tikzmark{preludeBegin}$
impl Memory { $\dots$ }
[[ Evaluation Functions ]]
fn get_input() -> Option<($T_{s_1}, \dots, T_{s_\ell}$)> {
  [[ Communicate with system ]]
}
fn emit(output: &($T_{s_1}, \dots, T_{s_n}$)) {
  [[ Communicate with system ]]
}
fn main() {
  let mut memory = Memory::new(); $\tikzmark{preludeEnd}$$\tikzmark{longest-pre}$
  let early_exit = prefix(&mem);
  if !early_exit {
    while let Some(input) = get_input() { $\tikzmark{longest}$$\tikzmark{loopBegin}$
	  mem.add_input(&input1);
      [[ Evaluation Logic ]]
    } $\tikzmark{loopEnd}$
  }
  postfix(&mem);
}
fn prefix(mem: &mut Memory) -> bool {
 if let Some(input) = get_input() {$\tikzmark{prefixBegin}$
   mem.add_input(&input);
	  [[ Evaluation Logic ]]
	} else {
     return true // Jump to Postfix.
	}
	[[ Repeat $\color[rgb]{0.6, 0.4, 0.08}\preflen$ times. ]] $\tikzmark{prefixEnd}$
	false // Continue with Monitor Loop.
}

fn postfix(mem &Memory) {
 [[ Evaluation Logic ]] $\tikzmark{postfixBegin}$
 [[ Repeat $\color[rgb]{0.6, 0.4, 0.08}\postlen$ times. ]] $\tikzmark{postfixEnd}$
}
$~$
$~$
$~$
$~$
$~$
$~$
\end{lstlisting}
\tikzset{
  line width=0.2pt
}
\vspace{-0.8cm}
\captionof{lstlisting}{Structure of the generated \rust code. The prelude is highlighted in orange, the monitor loop in blue, the execution prefix in green, and the execution postfix in violet.}\label{fig:monitorstructure}
\end{figure}

\Cref{fig:monitorstructure} illustrates the abstract structure of the generated Rust code.
The first portion of the code is the \emph{prelude} containing declaration for data structures and I/O functions.
Most notably, the \lstinline{Memory} struct represents the monitor's working memory and is of a fixed size.
For this, it utilizes the memory analysis from \Cref{sec:static:analyses}.
Note also, that the \lstinline{get_input} function returns an optional value: either it contains new input data or it indicates that the system terminated.

The \lstinline{main} function is the entry point of the monitor. 
It allocates the working memory and transitions to the prefix.
Here, the monitor code contains a static repetition of code checking for a new input, and evaluating all streams.
In the evaluation, stream access are either translated into immediate access to memory or substituted by constant default values.
The \lstinline{prefix} function returns a boolean flag indicating whether the system terminated before the prefix was completed.
This prompts the \lstinline{main} function to jump to the postfix immediately.
Otherwise, the main monitor loop begins following the same scheme of the prefix:  retrieve new input values, commit them to memory, evaluate streams, repeat until the system terminates.
Note that all stream accesses during the monitor loop translate to accesses to the working memory.
Lastly, the \lstinline{main} function triggers the computation of the postfix.
The structure is similar to the prefix except that it does not check for new input values.

The evaluation logic for streams is a straight-forward translation of the \lola specification as conditional statements, constants, and arithmetic functions are syntactically and semantically almost identical in \lola and Rust.
Only stream accesses requires special attention as they boil down to accesses to the \lstinline{Memory} struct.
Lastly, the compilation has to order the evaluation of streams to comply with the evaluation order from \Cref{sec:static:analyses}.
Streams in the same evaluation can be ordered arbitrarily or parallelized.
The latter leads to a significant runtime overhead and only pays off if computational complexity of the stream expressions is sufficiently high.

\subsubsection{Verification}


The compilation injects verification annotations in the Rust code to enable the Prusti~\cite{prusti} plugin of the Viper framework~\cite{viper} to verify functional correctness of the monitor.
The major challenge here is to verify that the finite memory available to the monitor suffices to accurately represent the infinite evaluation model of \lola.
The compilation achieves this by introducing a dynamically growing list of values, the ghost memory.
With this, the correctness proof proceeds in two steps.  
First, whenever the monitor receives or computes a new value, it commits it both to the ghost memory and the working memory.
Here, the working memory acts as a ring buffer: as soon as its capacity is reached, the addition of a new value overwrites and thus evicts the oldest value.
Therefore, the working memory is an excerpt of the ghost memory and thus the evaluation model.
Ergo, the computation of new values is valid with respect to the evaluation model because memory accesses yield the same values as the evaluation model would.
The newly computed, correct values, are then added into the memory, concluding the inductive proof of memory compliance.

Secondly, the verdict of a trigger in the theoretical model needs to be equivalent to the concrete monitor realization.
This amounts to proving that the trigger condition was translated properly.
Here, memory accesses are particularly interesting, because the theoretical computation uses entries of the ghost memory whereas the realization accesses the working memory only.
The agreement of the ghost memory and the working memory for the respective excerpts conclude this proof.

Note that for the monitor the ghost memory is write-only, whereas the verification procedure ``reads'' it, \ie, it refers to its theoretical values.
The evaluation logic of the monitor uses data from the system to compute output values.
Different components of the verification than access these values either directly or over the ghost memory (GM).
Clearly, the information flow is unidirectional: information flows from the monitor to the verification but not vice versa.
As a result, upon successful verification, the ghost memory can safely be dissected from the realization.

\subsubsection{Conclusion} 
Not only does the compilation from \lola to Rust produce performant runtime monitors, the injection of verification annotations answers the question \textit{``Quis custodiet ipsos custodes?}\footnote{\textit{``Who will guard the guards themselves?''}}'' rendering it an important step into the direction of verified runtime monitor.
The applicability of \lola for CPS is limited to high-level properties where neither asynchrony, nor real-time play a significant role.
Further research in this direction especially relating to \rtlola can significantly increase the value and practical relevance of the compilation.

\section{Integration and Post-Mortem}\label{sec:integration}

A key task when integrating a monitor into a CPS is finding a suitable spot in the system architecture.
Improper placement can lead to ineffective monitoring or negative interference, jeopardizing the safety of the entire system.

\subsection{Integration}

The integration is considerably easier when the architecture is not yet fully determined.
When adding the monitor retroactively, only minimal changes to the architecture are possible.
Larger changes would render previous tests void since the additional component might physically change the system, \eg in terms of weight distribution and power consumption, or logically offset the timing of other components.
Consider, for example, a monitor that relies on dedicated messages from the controller as input data source.
If the processor of the controller was already almost fully utilized, the additional communication leads to the controller missing deadlines.
This can lead to safety hazards, as timing is critical for the safe operation of a CPS.
Taking the placement of the monitor into account early on increases the degree of freedom, which helps avoid such problems.

The amount of interference the monitor imposes on the system also depends on the method of instrumentation.
Non-intrusive instrumentation such as bus snooping grants the monitor access to data without affecting other modules.
The effectiveness of this approach hinges on the amount of data available on the bus.

\begin{figure}[t]
\begin{subfigure}{.5\textwidth}
\centering
\includegraphics[width=\textwidth]{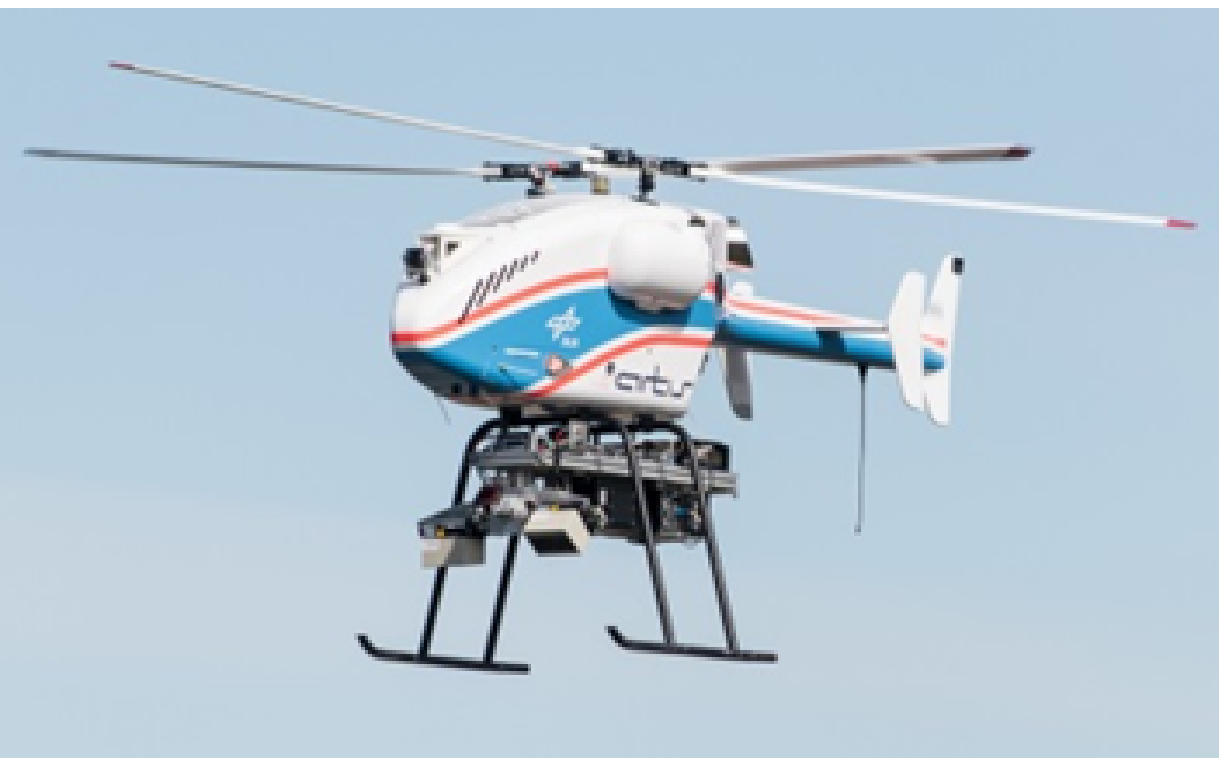}
\caption{An image of the superARTIS aircraft.}
\label{fig:superartis}
\end{subfigure}
\hfill
\begin{subfigure}{.45\textwidth}
\definecolor{mydarkblue}{rgb}{0.412,0.475,0.957}
\tikzstyle{state} = [circle, draw, text width=.5cm, text centered, minimum height=.5cm, minimum width=.5cm]
\tikzstyle{box} = [rectangle, draw, text centered, minimum height=1cm, minimum width=1cm, fill=white]
\tikzstyle{astate} = [circle, draw, text width=1.5cm, text centered, minimum height=2cm, minimum width=2cm]
\tikzstyle{smallcomponent} = [rectangle, draw, text width=6.5em, text centered, minimum height=3em, fill=white,font=\large]
\tikzstyle{component} = [rectangle, draw, text width=8em, text centered, minimum height=3em, fill=white]
\tikzstyle{smcomponent} = [component, minimum width=3cm, minimum height=2.5cm]
\tikzstyle{transition} = [draw, -stealth']
\tikzstyle{line} = [draw]
\tikzstyle{signal} = [draw, -latex']
\tikzstyle{config} = [densely dotted]

\colorlet{eventcolor}{green!50!black}
\colorlet{periodiccolor}{blue!50!black}
\tikzstyle{event} = [draw=eventcolor, thin, fill=white,fill opacity=.3, pattern=north west lines, pattern color=eventcolor]
\tikzstyle{periodic} = [draw=periodiccolor, thin, fill opacity=.3, pattern=north east lines, pattern color=periodiccolor]
\tikzstyle{signalname} = [text width=10em, minimum height=2em]
\tikzstyle{nameright} = [signalname, align=left]
\tikzstyle{namecenter} = [minimum height=2em, align=center, rotate=65]
\tikzstyle{nameleft} = [signalname, align=right]

\centering
\begin{tikzpicture}[scale=0.5, every node/.style={transform shape}]
  
    \node [smallcomponent, text width=1.8cm] at (2.8,2.1)  (dots) {\textsc{\dots}};
    \node [smallcomponent] at (1.4,0.9)  (imu) {\textsc{IMU}};
    \node [smallcomponent] at (0,2.1) (gnss) {\textsc{GNSS}};
    \node [smallcomponent] at (-1.4,0.9) (lidar) {\textsc{Lidar}};
    \node [smallcomponent] at (-2.8,2.1) (camera) {\textsc{Camera}};
    \node [box, font=\large] at (1.5,-1.7) (alg) {\textsc{PositionAlgorithm}};
    \node [box,minimum width=4cm, minimum height=3em,font=\large] at (0,-3.4) (logger) {\textsc{Logger}};
    \node [box,draw=monitorblue, minimum height=3em,thick, font=\large] at (-1.5,-5.1) (mon) {\textcolor{monitorblue}{\textsc{Monitor}}};
    \node [box,minimum height=3em, font=\large] at (1.5,-5.1) (disk) {\textsc{HardDisk}};

	\path [line,line width=0.5mm] (-3.5,0) -- (3.5,0);
	\path [signal] (dots) -- (2.8,0);
	\path [signal] (imu) -- (1.4,0);
	\path [signal] (gnss) -- (0,0);
	\path [signal] (lidar) -- (-1.4,0);
	\path [signal] (camera) -- (-2.8,0);
	\path [signal,line width=0.5mm] (0,0) -- (0,-0.5) -- (-1.5,-0.5) -- (-1.5,-2.9);
	\path [signal,line width=0.5mm] (0,-0.5) -- (1.5,-0.5) -- (alg);
	\path [signal,line width=0.5mm] (alg) -- (1.5,-2.9);
	\path [signal,line width=0.5mm] (1.5,-3.9) -- (disk);
	\path [signal,line width=0.5mm, color=monitorblue] (-1.5,-3.9) -- (mon);
    
  \end{tikzpicture}
  
\caption{A schematic of the superARTIS system architecture.}
\label{fig:systemarchitecture}
\end{subfigure}
\caption{Information on the superARTIS aircraft of the German Aerospace Center.}
\label{fig:superartisplusarch}
\end{figure}

Consider, for example, the system architecture of the autonomous aircraft superARTIS of the German Aerospace Center (DLR) depicted in \Cref{fig:superartisplusarch}.
When integrating a monitor for the optical navigation rail into the aircraft~\cite{rtlolacavindustrial}, the monitor was placed near the logging component.
By design, the logger had access to all relevant data.
This enabled monitoring of properties from the entire spectrum:  The specification contained single-sensor validation, cross-validation, and geo-fence compliance checks. 
Note that in this particular case, the utilization of the logger was low.
This allowed it to forward the information from the bus to the monitor.
In a scenario where performance is critical, snooping is the preferable option.

\subsection{Post-Mortem Analysis}

After termination of a flight, the post-mortem analysis allows for assessing the performance of the system and finding culprits for errors.
The analysis relies on data recorded during the mission; a full record enables a perfect reconstruction of the execution from the perspective of the system.
Resource restrictions, however, limit the amount of data, so full records are often not an option.
Thus, data needs to be filtered and aggregated rigorously.

A main task of the monitor is exactly this: refining input data by filtering and aggregation to obtain an accurate assessment of the system state.
Based on this assessment, the monitor determines whether a property is violated.
While this binary verdict is the major output of the monitor, the intermediate results provide valuable insight into the evolution of the system over time.
Hence, logging this data can improve the post-mortem analysis and alleviates the need to filter and aggregate data in another component as well.

%
%
%
%
%

\section{Bibliographic Remarks}\label{sec:literature}

Early work on runtime monitoring was mainly based on temporal logics~\cite{Drusinsky:2000:TRA:645880.672089,Lee99runtimeassurance, Finkbeiner+Sipma/01/Checking,ltl,Kupferman:2001:MCS:569028.569032,Havelund:2002:SMS:646486.694486}.
Their notion of time was limited to discrete time, leading to the development of real-time logics like \stl~\cite{STL} and \mtl~\cite{mtl}.
Not only do algorithms for runtime monitoring exist for these logics~\cite{RobustMonSTL,monitoringSTL,Basin:2015:MMF:2772377.2699444,aerial}, there is also work realizing it on an \fpga~\cite{stl2fpga}.
The \textsc{R2U2}~\cite{rtutjournal,rtutrv} tool in particular implements \mtl monitors on \fpga while allowing for future-time specifications.
Further, there are approaches for generating verified monitors for logics~\cite{metriccompiler,metriccompiler2}.

Apart from these temporal logics, there are other specification languages specifically for CPS such as differential dynamic logic~\cite{ddl}.
The ModelPlex~\cite{modelplex} framework translates such a specification into several verified components monitoring both the environment \wrt the assumed model and the controller decisions.

Other approaches --- such as the one presented in this paper --- completely forgo logics.
Similar to the compiler from \lola to Rust, there is a verified compiler for synchronous Lustre~\cite{lustrecompiler} programs to C code.
Moreover, the \copilot~\cite{copilot,copilotembedded} toolchain is based on a functional, declarative, stream language with real-time capabilities.
\copilot enables the verification of generated monitors using the \textsc{cbmc} model checker~\cite{cbmc}.
As opposed to the verification with \viper, their verification is limited to the absence of various arithmetic errors, lacking functional correctness.

In terms of integration, both the \textsc{R2U2}~\cite{rtutjournal} and the \copilot~\cite{copilotembedded} tool were successfully integrated into an aircraft.

\section{Conclusion}\label{sec:conclusion}
In this paper, I provided an overview of recent work on the development of a runtime monitor for cyber-physical systems from design to integration.
The process can roughly be divided into three phases.
In the specification phase, specifiers transform natural language descriptions of properties into a suitable formal specification language.
Such properties range from low-level properties validating a single sensor to high-level properties overseen the quality of the entire mission.
Type checking and validation based on log data from previous or simulated missions increase confidence in the specification.
The compilation phase transforms the specification into an executable software or hardware artifact, potentially injecting annotations to enable static verification of the monitor.
This process can help increase the effectiveness of the monitor, which directly translates into safer systems.

\section*{Acknowledgements}
This paper is based on a tutorial at the 20th International Conference on Runtime Verification.
The work summarized in this paper is based on several earlier publications~\cite{rtlolaarxiv,rtlolacavindustrial,rtlolacavtoolpaper,lolatorust,fpgalola} and I am grateful to all my co-authors.
I especially would also like to thank Jan Baumeister and Bernd Finkbeiner for providing valuable feedback and comments.

\bibliographystyle{splncs04}
\bibliography{bibliography}

\end{document}